\documentclass[prl,aps,tightenlines,twocolumn,nofootinbib]{revtex4}
\usepackage{graphicx}

\newcommand{\tm}{T_{\text{max}}}
\newcommand{\tr}{T_{\text{reh}}}
\newcommand{\mf}{m_{\phi}}
\newcommand{\beq}{\begin{equation}}
\newcommand{\eeq}{\end{equation}}
\newcommand{\beqy}{\begin{eqnarray}}
\newcommand{\eeqy}{\end{eqnarray}} 

\newcommand{\Gp}{\Gamma_{\phi}}

\newcommand{\rp}{\rho_{\phi}}
\newcommand{\rr}{\rho_{\text R}}

\newcommand{\hg}{\text{Hagedorn}}

\begin{document}

\title{A $\hg$ temperature in post-inflationary dynamics?}

\author{Prashanth Jaikumar and Anupam Mazumdar}
\affiliation{Physics Department, McGill University, 3600 University Street, 
Montr\'eal H3A 2T8, Canada }
\date{\today}

\begin{abstract}
We consider the possibility of regarding the maximum temperature $\tm$
of a plasma of inflaton decay products as a Hagedorn temperature
connected with the transition from an inflaton-dominated epoch to a 
radiation-dominated universe. Employing a ``bottom-up'' approach to 
inflaton decay, we show that the limiting temperature 
$T_{\rm H}\sim m_{\phi}$, as obtained from a numerical analysis of 
the evolution equations, can also be understood theoretically in a 
Hagedorn picture of the phase transition.
\end{abstract}

\maketitle
The concept of Hagedorn temperature~\cite{hagedorn65} in statistical 
systems is commonly invoked as a scale that signals the onset of a phase 
transition~\cite{Atick88}. While it does not directly relate to 
the usual thermodynamic definition of temperature, it can be regarded 
as a limiting temperature where the population of energy states begins 
to exceed the Boltzmann weight, implying that the original degrees of 
freedom strongly overlap in momentum space and lose their identity. 
One example of this is the melting of hadronic and glueball states in 
relativistic heavy-ion collisions, where the Hagedorn temperature 
$T_{\text{H}}$ can be derived from the Bag model~\cite{Chodos74} or a stringy 
model~\cite{Huang70} of hadrons, and can be used as an estimate of 
the deconfinement temperature, though the exact equivalence of the 
two is hard to establish, at least in four-dimensional gauge 
theories~\cite{Sundborg}. However, the idea of a number density of 
states that grows exponentially with temperature is quite natural 
coming from a theory with confining forces (like Regge theory), 
and may be useful when the exact nature of these microscopic forces 
is far from clear. A closely related example is Hagedorn-type behavior 
in some string theories at weak coupling~\cite{Atick88}. In the context 
of early string dominated cosmology, a Hagedorn-type phase could be  
associated with open strings attached to D-branes which has recently been 
shown to provide a negative pressure necessary to drive inflation 
at high temperatures near the string scale~\cite{Kogan}. This is 
accompanied by an exponential growth in the number of string modes 
with energy.

In this letter, we explore the possibility to have a Hagedorn limit 
of a quite different nature in cosmology, namely, in post-inflationary 
dynamics that culminate in reheating of the universe. Inflation
is a well known contender for making the Universe flat, besides generating 
adiabatic density perturbations with an almost flat spectrum~\cite{Liddle}. 
However, inflation must end in order to provide a thermal bath of
radiation in order to facilitate synthesis of light nuclei. The 
standard  paradigm is that the inflaton field, $\phi$, enters 
the oscillatory phase and decays perturbatively into radiation quanta, 
which subsequently thermalize completely through interactions, yielding 
a reheat temperature $\tr$, when the universe is completely radiation 
dominated~\cite{Turner}. It has also been proposed that an initial 
non-perturbative decay of the inflaton, dubbed as preheating, is a
plausible scenario~\cite{Robert}. Nevertheless there has to be a  
subsequent stage of thermalization after preheating quite similar to 
perturbative reheating~\cite{Sarkar}. Yet another interesting 
possibility is to have a non-topological defect formation from the 
fragmentation of the inflaton condensate, such as $Q$-balls~\cite{Mazumdar},
where reheating occurs via surface evaporation of $Q$-balls.

Irrespective of a particular scenario the final reheating temperature 
plays a very crucial role in the early universe, for e.g. see~\cite{Kari}. 
The reheat temperature can be constrained by bounds coming from thermal 
and non-thermal gravitino production during reheating/preheating in 
supersymmetric theories~\cite{Nanopoulos}. Typically, $\tr<10^{9}$ 
GeV is required, for gravitino mass $m_{3/2}\sim 100$~GeV, in order to 
avoid constraints coming from successful predictions of Big Bang 
Nucleosynthesis (BBN), for e.g. see \cite{Sarkarrep}.

The reheat temperature is not, however, the maximum temperature of 
a thermal plasma, which may be two orders of magnitude higher 
(but less than $m_{\phi}$)~\cite{Turner,Riotto}. Physically, this maximum 
temperature, denoted by $\tm$, arises due to the competition between 
early entropy production within light relativistic degrees of freedom 
at the earliest stages of inflaton decay and continued expansion of the 
universe which dilutes the energy density. The main objective of this 
work is to show that $\tm$ may be regarded as a limiting temperature, 
if we trace the inflaton decay process backwards in time, up to 
a point when coherent inflaton production from the radiation 
fields begins. We take into account thermal factors that capture 
the essential feature of Bose enhancement when the inflaton 
population builds up. We support our theoretical conjecture 
about this limiting temperature as a Hagedorn temperature with 
a numerical analysis of the evolution equations for matter 
(inflaton) and radiation energy densities run backwards in time.

After inflation ends, the energy density stored in
coherent oscillations of the inflaton field is transferred to 
radiation quanta over a period of time. Several features of this 
transition, such as thermalization~\cite{Sarkar}, associated 
heavy-particle production~\cite{Riotto}, and non-perturbative 
processes (preheating) have been widely investigated~\cite{Robert,Mazumdar}.

However, little is known about the dynamical process that results in 
inflaton decay. We may invert this problem, and ask the question 
as to how the inflaton field may be generated coherently from the 
radiation fields? Although the coupling between the two is small, 
our view is that Bose enhancement of the inflaton  in the final 
state can lead to coherence, similar to a condensate effect. 
Incorporating this feature in the Boltzmann equations that describe 
the perturbative decay of the inflaton is not hard. We may run the 
system of equations (see below) backwards in time (decreasing scale 
factor) to study how the energy densities evolve. We show that an 
exponential enhancement of the inflaton density can overcome the 
Boltzmann weight, very similar to the Hagedorn picture. The 
temperature at this instant is the limiting temperature, beyond 
which the radiation energy density drops rapidly to zero. This may 
be interpreted as $\tm$, the maximum or limiting temperature. Thus, 
we have an alternate physical description of the maximum temperature, 
and we will quantify it both theoretically and numerically in 
following.

We consider the interaction $L_{int}=g\phi{\bar f}f$ where $f$ 
denotes a fermion~\footnote{For Standard model quarks or leptons, 
such an interaction cannot be written down since it would not 
be a $SU(2)$ gauge invariant monomial. For the sake of exemplification, we 
assume that these fermions are pure singlet.}. The evolution 
equation for the inflaton energy density $\rho_{\phi}$ is given by
\begin{eqnarray}
\label{phieq}
\dot{\rho_{\phi}}+3H\rho_{\phi}&=&\Gp\rho_{\phi}(1+n_B(\mf))\,,\\
\Gp&=&\frac{g^2\mf}{32\pi}=\alpha_{\phi}\mf\,,
\end{eqnarray}
where $H$ is the Hubble expansion rate and $\Gamma_{\phi}$ is the
inflaton decay rate. Note that the source term for $\rho_{\phi}$ is
enhanced by the Bose factor $(1+n_B(\mf))=1/(1-{\rm e}^{-\mf/T})$,
since the inflaton is a scalar field. Since the expression for $\Gp$
is derived for zero total momentum (center-of-mass frame), the $p=0$
mode of the inflaton is being populated. $\rho_{\phi}$ as defined here
is the {\it thermal} energy density \beq
\rho_{\phi}=\frac{\mf^4}{({\rm e}^{\mf/T} -1)}\,.  \eeq The thermal
factor comes out naturally when one considers fermions described by a
thermal distribution fusing to form the inflaton.  The analogous
equation for $\rr$, the energy density of radiation is \beq
\label{radeq}
\dot{\rr}+4H\rr=-\Gp\rp(1+n_B(\mf))\,.
\eeq
We evolve eqns. (\ref{phieq}) and (\ref{radeq}) numerically, for 
which purpose it is convenient to define rescaled quantities as
\beq
x=a/\mf,\quad \Phi=\rho_{\phi} a^3,\quad R=\rr a^4/\mf\,.
\eeq  
where $a$ is the scale factor of the Friedman-Robertson-Walker metric. 
In terms of rescaled quantities, the evolution equations take the simple form
\beqy
\label{eveq}
\biggl(\frac{d\Phi}{dx}\biggr)&=&c\frac{x\Phi}{\sqrt{x\Phi+R}}(1+n_B(x,R)),\\
\biggl(\frac{dR}{dx}\biggr)&=&-c\frac{x^2\Phi}{\sqrt{x\Phi+R}}(1+n_B(x,R))\,.
\eeqy
where $c=\sqrt{\frac{3}{8\pi}}\alpha_{\phi}M_{Pl}\mf^{5/2}$. These 
equations are similar in form to those in~\cite{Riotto}, but the 
rescaling is different since we wish to study the build-up rather than 
the decay of the inflaton, for which time runs backwards. 
Furthermore, we include the Bose enhancement factor that 
was not required in~\cite{Riotto}. For the numerical analysis, we 
choose $\mf=10^{13}$ GeV, $\tr\sim 10^9$ GeV. The traditional 
expression for the reheat temperature corresponding to a thermal 
bath dominated solely by relativistic particles is given by~\cite{Turner}
\beq
\label{rh}
\tr=0.2\biggl(\frac{200}{g_*}\biggr)^{1/4}\sqrt{\alpha_{\phi}\mf M_{Pl}}\,,
\eeq
where $g_*$ accounts for the relativistic degrees of freedom in the 
plasma. Using this relation, and with $g_*=200$, we obtain 
$\alpha_{\phi}=2.3\times 10^{-13}$ and $c=8.7\times 10^{37.5}$. 
We begin the evolution from $\tr\sim 10^9$ GeV, which fixes the 
radiation energy density through the Stefan-Boltzmann law
\beq
\rr=\biggl(\frac{g_*\pi^2}{30}\biggr)^{1/4}T^4\approx 
2.85\times 10^{36}~{\rm GeV}^4
\eeq
However note that in our case there is no inflaton oscillation dominated
phase and radiation is the only component we start with. In this respect
the traditional notion of reheat temperature, which applies to a largest
temperature of a thermal bath only dominated by relativistic particles, 
does not hold in this scenario. Nevertheless the numerical significance of
eqn.~(\ref{rh}) continues to be important as an estimation of a temperature
of a relativistic bath when the entire inflaton quanta has been converted into 
radiation.

\begin{figure}[h!]
\begin{center}
\includegraphics[scale=0.4,angle=270,clip=]{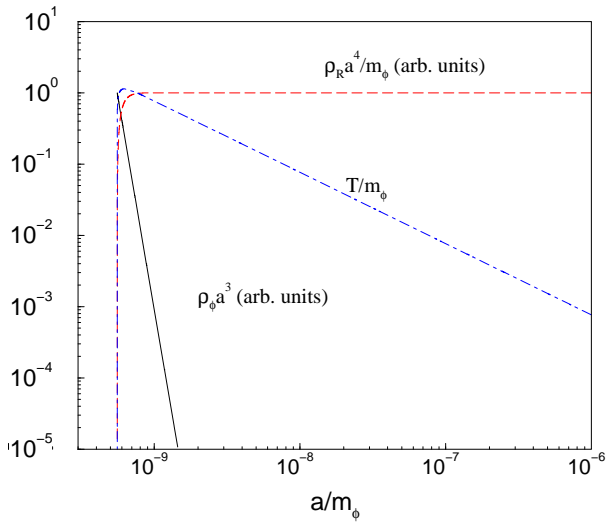}
\end{center}
\caption{Evolution (backwards in time or with decreasing scale factor) 
of the inflaton and radiation energy densities. The value of the 
limiting temperature is at the peak $\tm\approx\mf$.}
\label{figps1}
\end{figure}

The qualitative features of the result of the numerical analysis are
independent of the choice of initial value of scale factor $a$,
although $x,\Phi, R$ are quantitatively dependent on it. We choose
$a=10^8~(x=10^{-5})$ at the beginning, and the results of the
evolution across nearly 4 e-foldings are displayed in
Fig.~\ref{figps1}.

While the radiation curve $R(x)$ (long dashed line) is remaining 
constant, the temperature $T\propto a^{-1}$, as expected for the 
radiation-dominated epoch. This is because both $\rr$ and $a$ are 
changing. As the inflaton energy density (solid line) builds up, the 
radiation energy density drops rapidly to zero. At this point, it is 
easy to see from eqn.(\ref{eveq}) that $\Phi(x)\propto x^3$, which 
is the usual initial condition obtained in standard evolution of 
the inflaton decay equation. Physically, our picture is quite 
different from that in~\cite{Riotto}, in that we can interpret 
$\tm$ as a limiting temperature, where the inflaton fields acquires 
a coherent nature, and the phase transition to an inflaton-dominated 
universe is underway. This interpretation, though novel, can be supported 
by a simple analytical estimate of $\tm$ which we presently derive. 

It is also interesting to observe that in our case the process of 
thermalization is actually occurring very close to $\tm$ in such a 
way that entropy density of a radiation dominated bath follows an 
adiabatic expansion law $sa^3={\rm constant}$, where $s$ is the entropy 
density. In this respect we are tempted to associate $\tm$ as an actual 
reheat temperature of the universe.

At the limiting temperature, when the inflaton energy density is
dominant, $\Gp\gg H$ holds, and we may neglect the $3H\dot{\rp}$ term
in eqn.(\ref{phieq}). Assuming the temperature to be approximately
constant over a small range of time, we may integrate the equation
from time $t_0$ to $t$ ($t_0<t$ in the absolute sense) to obtain \beq
{\rm ln}\biggl(\frac{n_{\phi}(t)}{n_{\phi}(t_0)}\biggr)=
\frac{\alpha_{\phi}\mf (t-t_0)}{1-{\rm e}^{-\mf/T}}\,, \eeq where
$n_{\phi}=\rp/\mf$. This equation governs the rate of increase in
the inflaton number density in a small interval $t-t_0$ near the
transition. This time interval must be of the order of (or slightly
less than) $1/\Gp\lesssim1/(\alpha_{\phi}\mf)$, because subsequently,
the inflaton density drops off much more rapidly than the
temperature. This means we may approximate as 

\beq n_{\phi}(t)\approx
n_{\phi}(t_0){\rm e}^{1/(1-{\rm e}^{-\mf/T})}\,.  
\eeq 

Now, since
$n_{\phi}(t_0)$ includes a statistical weight factor, ${\rm
e}^{-\mf/T}$, (we assume $T\leq\mf$ so that it is approximately a
Boltzmann weight), the growth in number density of inflatons begins to
overwhelm the exponential suppression from the statistical weight when

\beq 
T\approx\mf(1-{\rm e}^{-\mf/T})\,, 
\eeq 

which is uniquely satisfied for $T\approx 0.74\mf$. We regard this as
a maximum or limiting temperature, reminiscent of the Hagedorn
temperature from hadronic physics. This is the main result of this
work. Beyond this temperature, the radiation fields overlap strongly
and populate the zero momentum mode of the inflaton at a rate that is
sufficient to overcome the thermal suppression. The universe then has
only an inflaton component to its energy density. We see that the
analytical estimate of $T_{\rm max}$ or $T_{\text{H}}$ is close to the
numerical value obtained from evolving the Boltzmann equations for
inflaton decay backwards in time. This numerical estimate is robust
and is not affected by inclusion of inverse reaction terms in the
collision kernel of the Boltzmann equations, which only become
significant at temperatures exceeding the limiting temperature.
In~\cite{Kolb:2003ke}, modulo extra assumptions on the gauge
interactions of the fermions and their coupling to the gauge fields,
it was argued that $T_{\rm max}$ could exceed $m_{\phi}$.  We find
that in the Hagedorn picture, including thermal masses for the
fermions, as in~\cite{Kolb:2003ke}, does not affect the bound $T_{\rm
max}\approx 0.74m_{\phi}$.

\vskip 0.1cm

In summary, we have shown that it is possible to interpret the maximum 
temperature reached in the evolution of inflaton decay as a Hagedorn 
temperature. Viewed backwards in time, or with decreasing size of the 
universe, at this temperature, which is close to $\mf$, coherent 
population of the zero momentum state of the inflaton takes place. 
The rate of population exceeds the thermal suppression factor, and 
it is not appropriate to use the evolution equations any longer, 
since a clear distinction between radiation and inflaton fields 
cannot be made. This is similar to the purported hadron gas to 
quark-gluon plasma (QGP) transition at finite temperature $T\sim m_{\pi}$, 
where the Hagedorn temperature signifies when the quark fields start to 
percolate through the overlapping hadrons. At this point, hadrons are no 
longer genuine physical degrees of freedom, and the deconfinement phase 
transition is approached. We have applied the same idea to the early 
stage of inflaton decay, and supported the conjecture with a numerical 
analysis of the Boltzmann equations.  
        
We find that the maximum temperature, or the corresponding Hagedorn
temperature of the post-inflationary universe is very close to the
inflaton mass, $\tm\approx 0.74m_{\phi}$.  We regard this is as an actual
reheat temperature of the universe because below the phase transition,
the subsequent evolution of radiation density obeys an adiabatic
expansion law with $T\propto a^{-1}$. We consider this result as having
an impact on particle-cosmology which we shall explore in a separate
publication.

We thank Jim Cline for important remarks. P. J. is
supported in part by the Natural Sciences and Engineering Research
Council of Canada and in part by the Fonds Nature et Technologies of
Qu\'ebec.  A. M. is a CITA national fellow.


\end{document}